# Scalable Growth of High Mobility Dirac Semimetal Cd$_3$As$_2$ Microbelts

Zhi-Gang Chen,[1*†] Cheng Zhang,[2,3†] Yichao Zou,[1] Enze Zhang,[2,3] Lei Yang,[1] Faxian Xiu,[2,3*] and Jin Zou[1,4*]

[1]Materials Engineering, The University of Queensland, Brisbane QLD 4072, Australia

[2]State Key Laboratory of Surface Physics and Department of Physics, Fudan University, Shanghai 200433, China

[3]Collaborative Innovation Center of Advanced Microstructures, Fudan University, Shanghai 200433, China

[4]Centre for Microscopy and Microanalysis, The University of Queensland, Brisbane QLD 4072, Australia

[†] These authors contributed equally to this work.

[*] Correspondence and requests for materials should be addressed to J. Z (j.zou@uq.edu.au), F. X. (E-mail: faxian@fudan.edu.cn), and Z. G. C. (z.chen1@uq.edu.au).

KEYWORDS: Cd$_3$As$_2$, Microbelts, Dirac semimetals, Weyl fermions

**Three dimensional (3D) Dirac semimetals are 3D analogue of graphene, which display Dirac points with linear dispersion in *k*-space, stabilized by crystal symmetry. $Cd_3As_2$ and $Na_3Bi$ were predicted to be 3D Dirac semimetals and were subsequently demonstrated by photoemission experiments. As unveiled by transport measurements, several exotic phases, such as Weyl semimetals, topological insulators, and topological superconductors, can be deduced by breaking time reversal or inversion symmetry. Here, we reported a facile and scalable chemical vapor deposition method to fabricate high-quality Dirac semimetal $Cd_3As_2$ microbelts, they have shown ultrahigh mobility up to $1.15 \times 10^5 \text{ cm}^2/\text{V s}$ and pronounced Shubnikov-de Haas oscillations. Such extraordinary features are attributed to the suppression of electron backscattering. This research opens a new avenue for the scalable fabrication of $Cd_3As_2$ materials towards exciting electronic applications of 3D Dirac semimetals.**

Owing to unique band structure and inherent exotic physical properties originated from Dirac fermions with linear band dispersion, Dirac materials, such as graphene[1] and topological insulators,[2] have been considered as a promising candidate for next-generation electronic and spintronic devices. As the emerging three-dimensional (3D) Dirac materials, $Cd_3As_2$[3] and $Na_3Bi$[4] have been theoretically predicted and subsequently verified by the angle-resolved photoemission spectroscopy (ARPES)[5-8] and electric transport measurments,[9-12] in which the Dirac nodes are developed via the point contact of conduction-valence bands with linear dispersion in all 3D of *k*-space. More importantly, 3D Dirac semimetals can be driven into various novel phases, such as Weyl semimetals,[13, 14] topological insulators,[15] axion insulators,[16, 17] and topological superconductors,[12] by breaking time reversal symmetry or inversion symmetry. Thus, 3D Dirac semimetal is a versatile platform for the systematic study of unusual quantum phase transitions between rich topological quantum states.[4]

Compared with the air-sensitive $Na_3Bi$, $Cd_3As_2$ tends to be much stable at room temperature with a high chemical stability against oxidation,[18] which is an ideal system for 3D Dirac materials. Soon after the first prediction in $Cd_3As_2$,[3] photoemission spectroscopy revealed a pair of 3D Dirac nodes located on the opposite sides of the Brillouin zone center (Γ point), which are protected by the $C_4$ rotational symmetry.[5] Transport measurements show that $Cd_3As_2$ exhibits ultrahigh mobility up to $9 \times 10^6$ $cm^2V^{-1}s^{-1}$ at 5 K,[9] a giant linear magnetoresistance,[10] and a nontrivial Berry phase[11] owing to the linear band dispersion and concomitant Dirac fermions. Very recently, a pressure-induced superconductivity was also identified in $Cd_3As_2$ crystals, making it an promising candidate for the topological superconductors.[12] All these studies relied on the high-quality crystals prepared from the time consuming Flux method, which has hindered the research expansion and future practical applications.

Here, we reported a new approach - facile and scalable chemical vapor deposition (CVD) method - to fabricate high-quality $Cd_3As_2$ microbelts. The individual single-crystal $Cd_3As_2$ microbelts exhibit ultrahigh mobility up to $1.15 \times 10^5$ cm$^2$/V s and pronounced Shubnikov-de Haas (SdH) oscillations, which are attributed to the suppression of electron backscattering. This study suggests that CVD is a powerful approach to fabricate high-quality $Cd_3As_2$ Dirac semimetals. The successful synthesis of $Cd_3As_2$ micro-sized crystals allows the demonstration of their groundbreaking physical properties, which in turn opens up potential electronic applications of 3D Dirac semimetals.

We grew $Cd_3As_2$ microbelts using $Cd_3As_2$ powders as the precursor in a horizontal tube furnace. After a typical growth procedure (see methods), high-quality needle-like crystals grown on silicon substrates are observable by naked-eyes. Figure 1a is a scanning electron microscopy (SEM) image that shows the typical morphology of as-grown microbelts, from which the belt-like crystals exhibit a length up to millimeter with smooth surfaces. The typical width and thickness are up to tens of micrometers. In order to determine their crystal structure, we carried out X-ray diffraction (XRD) analysis on a single microbelt with the belt surface normal to the XRD beam. Figure 1b shows the typical XRD pattern, in which a series of sharp peaks can be indexed to $\{n,n,2n\}$ (where n is an integer), further verifying that the longitudinal shining surfaces are $\{112\}$ planes with $<1\bar{1}0>$ axial directions. In order to identify other surfaces, we viewed a microbelt along its growth direction. Figure 1c is a typical cross-section image and the inset displays its longitudinal image. Based on Figure 1c, we can measure the angle between adjacent facets as $120^0$, suggesting that these four surfaces are equivalents as they are all $\{112\}$ surfaces. To better understand this relationship, we draw an atomic model viewed along the $[1\bar{1}0]$ direction, as shown in Figure 1d. As marked by the arrows, 4 equivalent $\{112\}$ surfaces can be observed. To further reveal the element distribution of Cd and As in this microbelt, we performed SEM element mapping and Figure 1e demonstrates the uniform distribution of Cd and As in the entire micorbelt.

To understand the structural characteristics of as-grown $Cd_3As_2$ microbelts, we conducted transmission electron microscopy (TEM) investigations. Figure 2a shows a typical TEM image of a $Cd_3As_2$ thin flake on a holey carbon film. Figure 2b and Figure 2c are the corresponding selected area electron diffraction pattern and high-resolution TEM images taken along the quasi [221] zone axis, respectively. Figure 2c evidences a perfect crystalline structure and can be indexed as the (112) plane. Figure 2d shows a typical EDS spectrum with an atomic ratio of Cd:As=3:2, which verifies their perfect stoichiometry in as-grown $Cd_3As_2$ crystals. Based on the analyses outlined above, Figure 2e illustrates the atomic model of the obtained $Cd_3As_2$ crystals, which is an antifluorite ($M_2X$) structure type and belongs to a space group of $I4_1/acd$,[19] in which ¼ of the 64 Cd sites in each unit cell are vacant in the ideal lattice.

To investigate the transport properties of as-grown $Cd_3As_2$ micorbelts, we fabricated a Hall bar device with standard six-terminal geometry, as schematically illustrated in Figure 3a. A constant current was applied along the $[1\bar{1}0]$ direction while the magnetic field was titled from the $[221]$ to $[1\bar{1}0]$ directions. Figure 3b shows the temperature dependence of longitudinal resistivity $\rho_{xx}$ at zero magnetic field for a typical $Cd_3As_2$ microbelt. The $\rho_{xx}$-T curve reveals a metallic behavior of our $Cd_3As_2$ arising from the semimetal band structure. Below 10 K, the curve is flat, extrapolating to a residual resistivity $\rho_{xx0} \approx 4\mu\Omega cm$. One of the most fascinating features of our $Cd_3As_2$ microbelts is the ultrahigh mobility deriving from the linear band dispersion. Figure 3c shows a Hall measurement from a typical $Cd_3As_2$ crystal, from which a high electron mobility of $\mu = 2 \times 10^4$ cm$^2$/V s at room temperature can be witnessed. In fact, most of our samples have a room-temperature mobility in the order of $10^4$ cm$^2$/V s. From the temperature dependence of carriers mobility (Figure 3d), a significant mobility increase (up to $1.15 \times 10^5$ cm$^2$/V s) is observed at 2.1K, which is comparable to most other reported values.[10, 20] Such a high mobility should be resulted from the suppression of electron backscattering.[9, 11] Our $Cd_3As_2$ microbelts also exhibits a relative lower carrier density of $n_e = 1.36 \times 10^{19}$ cm$^{-3}$ at 2.1 K and experiences a negligible change with temperature, as shown in Figure 3d.

To determine the Fermi surface (FS) of our $Cd_3As_2$, we carried out the magnetotransport measurements using a physical properties measurement system (PPMS, B≤9T). Figure 4a shows $\rho_{xx}$ under the different magnetic fields (B≤ 9T), which show pronounced SdH oscillations. The magnetoresistivity (MR) is defined by MR=[$\rho_{xx}$(B)- $\rho_{xx}$(0T)]/$\rho_{xx}$(0T) ×100%. Within the testing temperature range, the MR shows no sign of saturation in high field (B>3T), which is consistent with the theoretical prediction.[3] At 2.1K, a MR = ~ 135% can be estimated.

To fundamentally understand the SdH oscillations, we displayed the oscillatory component of $\Delta\rho_{xx}$ versus 1/B at various temperatures after subtracting a smooth background. The results are shown in Figure 4b, in which a single oscillation frequency F = 32.15 T is identified from the fast Fourier transform spectra, corresponding to a periodicity of $\Delta(1/B) = 0.0311 \text{ T}^{-1}$. According to the Onsager relation:[20] $F = (\phi_0/2\pi^2)S_F$, where $\phi_0 = h/2e$, the cross-sectional area of the FS normal to the field is $S_F = 2.94 \times 10^{-4} \text{ Å}^{-2}$. By assuming a circular cross-section, the Fermi vector $k_F = 0.0309 \text{ Å}^{-1}$ can be extracted. This value is slight smaller than the recent ARPES experiment, which shows that the FS of $Cd_3As_2$ consists of two tiny ellipsoids or almost sphere with $k_F = 0.04 \text{ Å}^{-1}$.[7]

The SdH oscillation amplitude as a function of temperature is also analyzed to clarify the nature of the carrier transport of our $Cd_3As_2$. The temperature-dependent amplitude $\Delta\sigma_{xx}$ can be described by $\Delta\sigma_{xx}(T)/\Delta\sigma_{xx}(0) = \lambda(T)/\sinh(\lambda(T))$,[21] and the thermal factor is given by $\lambda(T) = 2\pi^2 k_B T m_{cycl}/(\hbar eB)$,[21] where $k_B$ is the Boltzmann's constant, $\hbar$ is the reduced plank constant, and $m_{cyc} = E_F/v_F^2$ is the cyclotron mass. By taking conductivity oscillation amplitude and performing the best fit to the $\Delta\sigma_{xx}(T)/\Delta\sigma_{xx}(0)$ equation, $m_{cyc} = 0.032 m_e$ can be extracted, as shown in Figure 4d. Using the equation $v_F = \hbar k_F/m_{cyc}$,[21] we can obtain the Fermi velocity of $v_F = 1.12 \times 10^6 \text{ m/s}$. This $v_F$ value is close to the ARPES result of $1.5 \times$ $10^6$ m/s.[6] Such a large Fermi velocity may explain the unusual high carrier mobility found in our Cd$_3$As$_2$ microbelts. According to $E_F = m_{cyc} \times v_F^2$,[21] $E_F = 228$ meV can be obtained, indicating that a band dispersion in our Cd$_3$As$_2$ is in prefect agreement with the previous scanning tunneling microscopy (STM) results.[8] All the key determined parameters, derived from the SdH oscillations at B≤ 9T, are summarized in Table 1.

The Berry's phase can be obtained from the Landau fan diagram (Figure 4e). According to the Lifshitz-Onsager quantization rule:[22] $S_F \frac{\hbar}{eB} = 2\pi\left(n + \frac{1}{2} - \frac{\phi_B}{2\pi}\right) = 2\pi(n + \gamma)$, the offset $\gamma$ in the Landau fan diagram gives the Berry phase $\phi_B$ by $\gamma = \frac{1}{2} - \frac{\phi_B}{2\pi}$, where $S_F$ is the cross-sectional area of the FS (corresponding to Landau level n), $\hbar$ is the reduced Planck's constant, and e is the elementary charge. For a nontrivial Berry phase, $\gamma$ should be 0 or 1. From the Landau fan diagram of Figure 4e, the intercept of our Cd$_3$As$_2$ microbelts is subtracted to be $-2.43 \times 10^{-4}$, which can be considered as close to zero. This intercept clearly reveals the nontrivial $\pi$ Berry phase and thus provides strong evidence for the existence of Dirac fermions in our Cd$_3$As$_2$ microbelts.[20] In fact, such a nontrivial $\pi$ Berry phase has been observed in the 2D graphene,[1, 23] bulk SrMnBi$_2$,[24] Rashba semiconductor BiTeI,[25] and bulk Cd$_3$As$_2$,[11] which indicates a Dirac system with linear dispersion.

In conclusion, we show a new fabrication approach to fabricate high-quality 3D Dirac semimetal Cd$_3$As$_2$ microbelts, evidenced by their demonstrated ultrahigh mobility and pronounced SdH oscillations measured from the single crystal Cd$_3$As$_2$ microbelts. The fundamental reason response to the observed appealing performance is attributed to the suppression of electron backscattering. This study indicates that CVD growth is a powerful approach to fabricate high-quality Dirac semimetals, such as $Cd_3As_2$ microbelts. The successful synthesis of high-quality $Cd_3As_2$ microbelts by such a new facile method and their demonstrated rich physical properties open up exciting electronic applications of 3D Dirac semimetals.

**Acknowledgements**

This work was supported by Australian Research Council, the China National Young 1000 Talent Plan, Shanghai Pujiang Talent Plan, National Natural Science Foundation of China (61322407, 11474058), and the Chinese National Science Fund for Talent Training in Basic Science (J1103204). ZGC thanks QLD government for a smart state future fellowship. The Australian Microscopy & Microanalysis Research Facility and the Queensland node of the Australian National Fabrication Facility are acknowledged for providing characterization facilities.

**Competing financial interests**

The authors declare no competing financial interests.

**FIGURES**

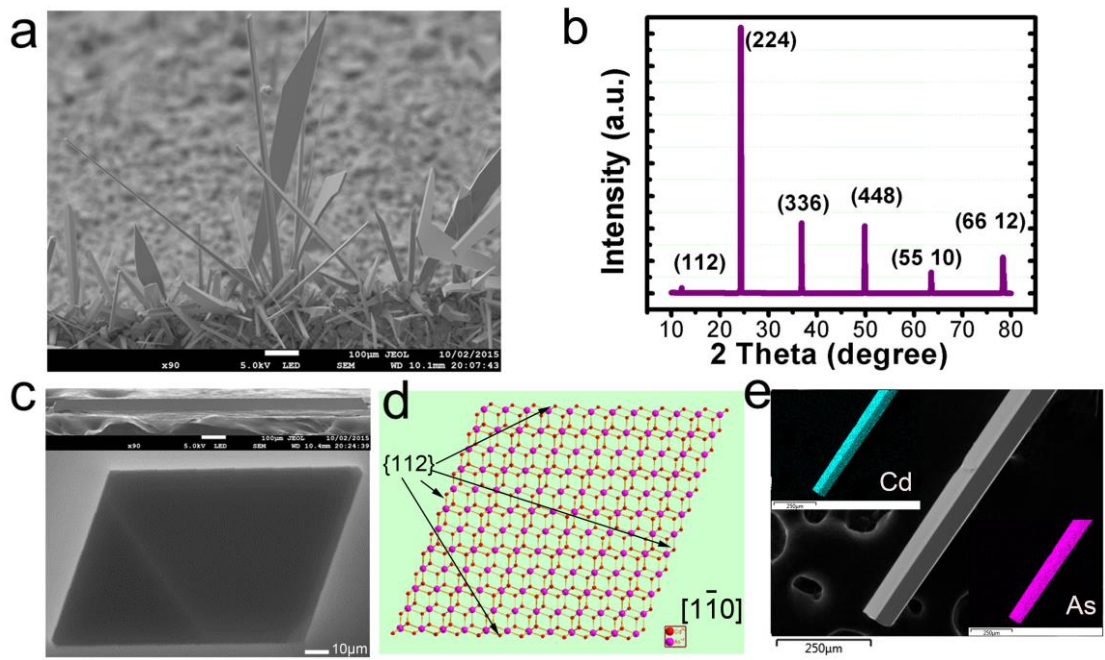

**Figure1| Structural and electrical properties of Cd$_3$As$_2$ microbelts. a,** A Typical SEM image. **b,** A X-ray diffraction pattern of the single crystal Cd$_3$As$_2$. The peak position shows that the sample surface is (112) plane. **c,** Cross-section view of a typical Cd$_3$As$_2$ microbelt and inset is longitudinal view image. d, Atomic plane $[1\bar{1}0]$, showing that the side walls are (112) planes. **e**, Cd and As EDS-SEM mapping of a typical Cd$_3$As$_2$ microbelt.

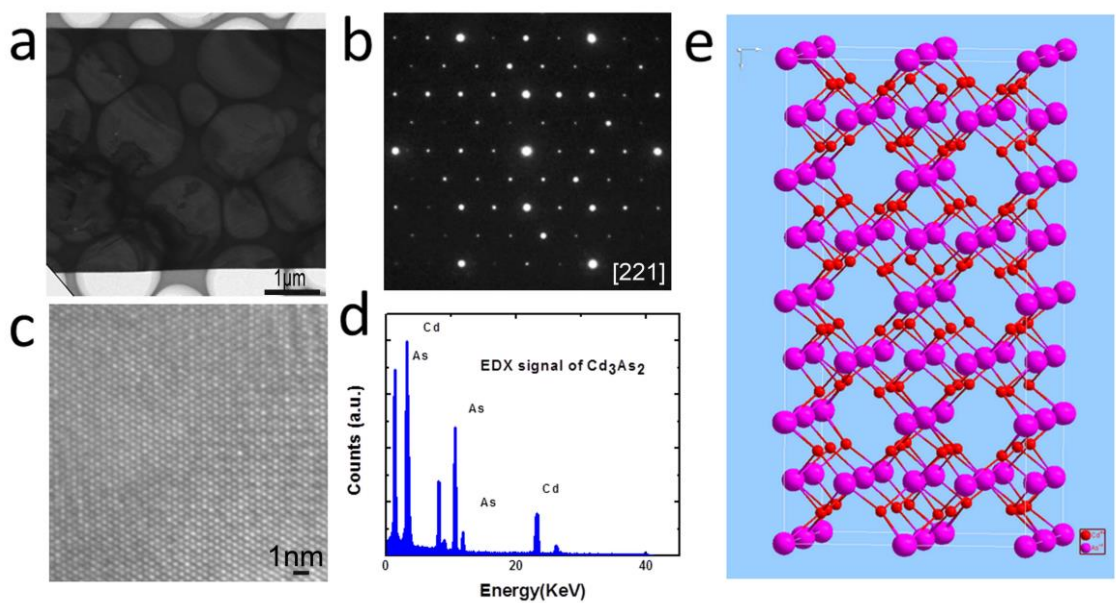

**Figure 2| Structural and electrical properties of Cd$_3$As$_2$ microbelts. a,** A typical TEM image of a Cd$_3$As$_2$ thin flake on a holey carbon grid taken along [221] direction, **b**, selected area diffraction pattern, **c**, High magnification TEM picture revealing a perfect crystalline structure. **d,** A typical EDS spectrum showing the atomic ratio of Cd:As=3:2. **e,** Crystal model at [010] axis.

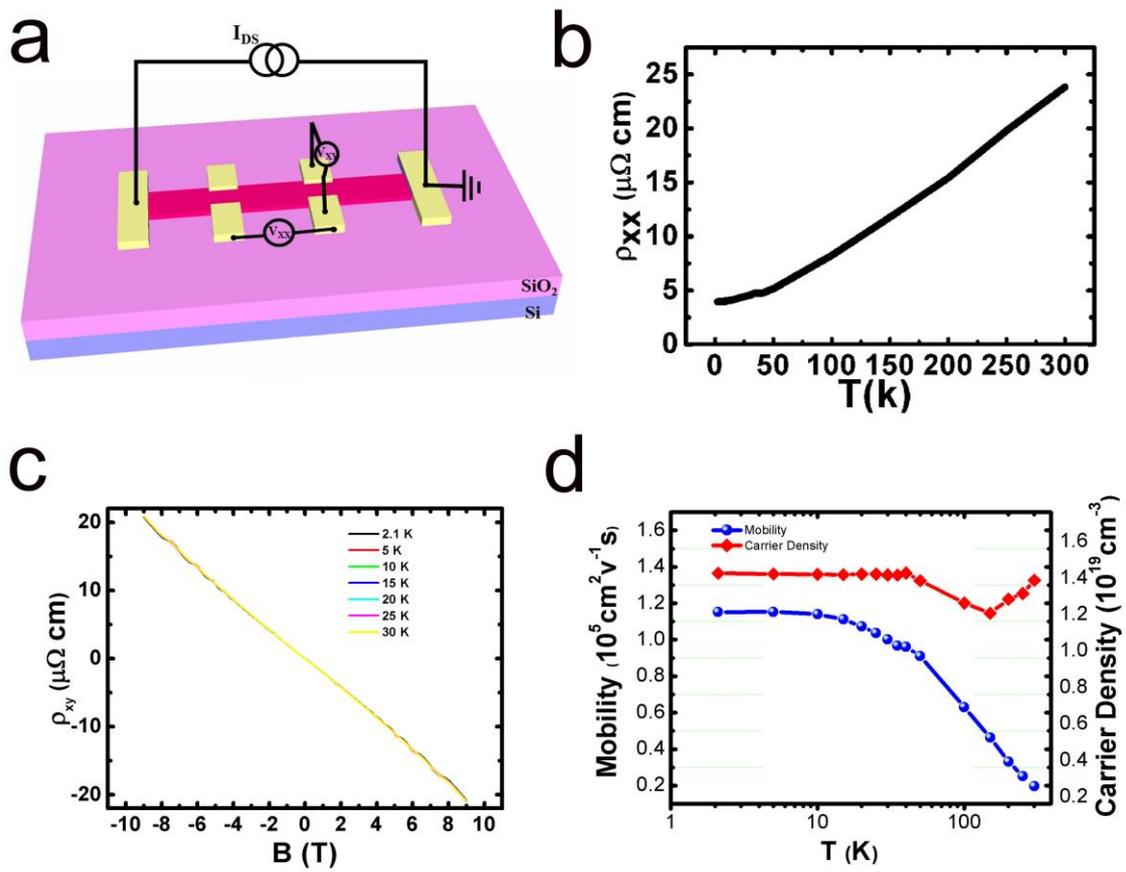

**Figure 3| Electrical properties of Cd$_3$As$_2$ microbelts.** a, A typical 6-terminal Hall bar geometry device with an applied current along [110] direction. **b,** The longitudinal resistivity as a function of temperature, showing a typical metallic behavior. **c**, The axial resistivity as a function of magnetic field at different temperatures. **d,** The temperature-dependent mobility and carrier density from 2.1 to 300 K. At 2.1 K, the mobility reaches $1.15 \times 10^5$ cm$^2$/V s.

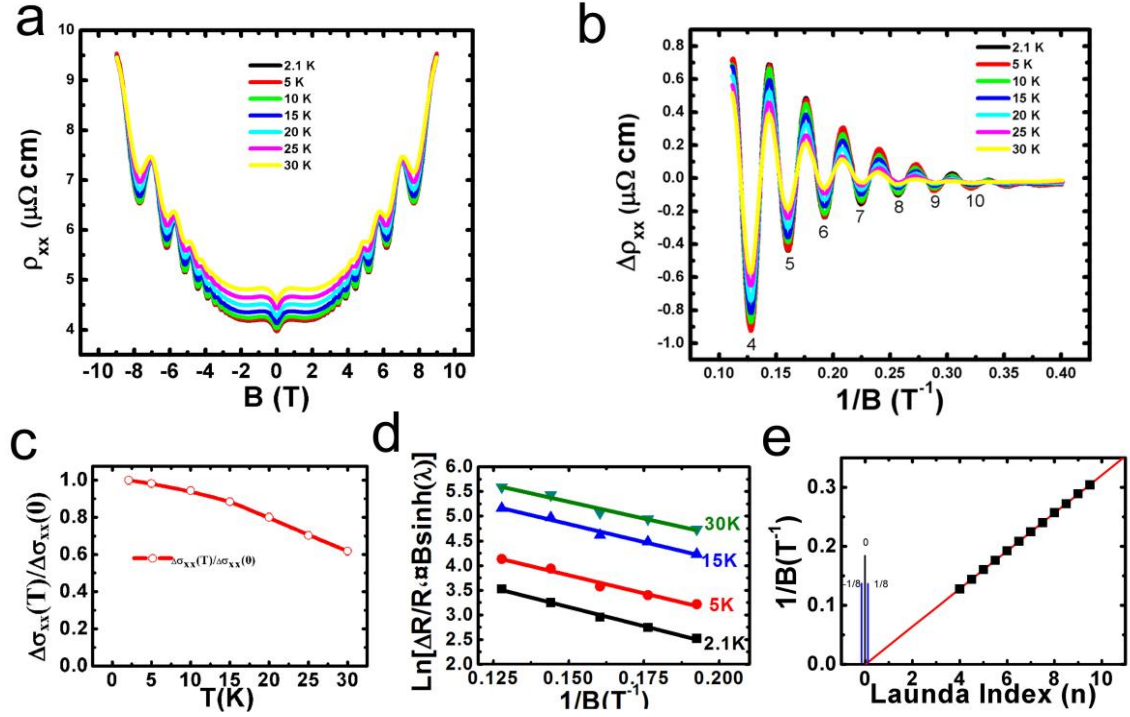

**Figure 4| Low magnetic field transport measurements (B≤9 T). a,** The longitudinal resistivity at different temperatures at θ=0°. The critical temperature is found to be 30 K, above which the oscillation is not observable. **b,** The oscillatory component Δρ$_{xx}$, extracted from ρ$_{xx}$ by subtracting a smooth background, as a function of 1/B at various temperature, **c,** Normalized conductivity amplitude versus temperature. The outcome can be fitted with the equation Δσ$_{xx}$(T))/(Δσ$_{xx}$(0)=λ(T))/(sinh(λ(T)) and the R-square is higher than 0.999 (the coefficient of multiple determination). **d,** The Dingle plot of the sample after an FFT process at θ=0° with 4 T magnetic field. **e,** Landau index plot of the Δρ$_{xx}$ peaks and valley from b. The intercept is between −1/8 and 1/8 for the sample.

**Table 1| Estimated parameters from the SdH oscillations (B≤9 T).**

| | $B_F(T)$ | $m^*$ | $S_F(Å^{-2})$ | $K_F(Å^{-1})$ | $\tau(s)$ | $V_f(m/s)$ | $L(nm)$ | $E_F(meV)$ |
|---|---|---|---|---|---|---|---|---|
| **Sample** | 31.14 | $0.032\, m_0$ | $2.98\times10^{-3}$ | 0.0308 | $1.11\times10^{-13}$ | $1.1155\times10^6$ | 124 | 228 |

# References

[1] K. S. Novoselov, A. K. Geim, S. V. Morozov, D. Jiang, M. I. Katsnelson, I. V. Grigorieva, S. V. Dubonos, A. A. Firsov, Nature 2005, 438, 197.

[2] H. J. Zhang, C. X. Liu, X. L. Qi, X. Dai, Z. Fang, S. C. Zhang, Nat. Phys. 2009, 5, 438.

[3] Z. J. Wang, H. M. Weng, Q. S. Wu, X. Dai, Z. Fang, Phys. Rev. B 2013, 88, 125427.

[4] Z. K. Liu, B. Zhou, Y. Zhang, Z. J. Wang, H. M. Weng, D. Prabhakaran, S. K. Mo, Z. X. Shen, Z. Fang, X. Dai, Z. Hussain, Y. L. Chen, Science 2014, 343, 864.

[5] Z. K. Liu, J. Jiang, B. Zhou, Z. J. Wang, Y. Zhang, H. M. Weng, D. Prabhakaran, S. K. Mo, H. Peng, P. Dudin, T. Kim, M. Hoesch, Z. Fang, X. Dai, Z. X. Shen, D. L. Feng, Z. Hussain, Y. L. Chen, Nat. Mater. 2014, 13, 677.

[6] M. Neupane, S.-Y. Xu, R. Sankar, N. Alidoust, G. Bian, C. Liu, I. Belopolski, T.-R. Chang, H.-T. Jeng, H. Lin, A. Bansil, F. Chou, M. Z. Hasan, Nat. Commun. 2014, 5, 3786.

[7] S. Borisenko, Q. Gibson, D. Evtushinsky, V. Zabolotnyy, B. Büchner, R. J. Cava, Phys. Rev. Lett. 2014, 113, 027603.

[8] S. Jeon, B. B. Zhou, A. Gyenis, B. E. Feldman, I. Kimchi, A. C. Potter, Q. D. Gibson, R. J. Cava, A. Vishwanath, A. Yazdani, Nat. Mater. 2014, 13, 851.

[9] T. Liang, Q. Gibson, M. N. Ali, M. Liu, R. J. Cava, N. P. Ong, Nat Mater 2015, 14, 280.

[10] Y. P. Junya Feng, Desheng Wu, Zhijun Wang, Hongming, J. L. Weng, Xi Dai, Zhong Fang, Youguo Shi, and Li Lu, arXiv:1405.6611 2014.

[11] A. Narayanan, M. D. Watson, S. F. Blake, Y. L. Chen, D. Prabhakaran, B. Yan, N. Bruyant, L. Drigo, I. I. Mazin, C. Felser, T. Kong, P. C. Canfield, A. I. Coldea, arXiv:1412.4105.

[12] L. P. He, Y. T. Jia, S. J. Zhang, X. C. Hong, S. Y. Li, arXiv:1502.02509 2014.

[13] A. A. Burkov, L. Balents, Phys. Rev. Lett. 2011, 107, 4.

[14] Junzhi Cao, S. Liang, C. Zhang, Y. Liu, J. Huang, Z. Jin, Z.-G. Chen, Z. Wang, Q. Wang, J. Zhao, S. Li, X. Dai, J. Zou, Z. Xia, L. Li, F. Xiu, arXiv:1412.0824.

[15] M. Z. Hasan, C. L. Kane, Rev. Mod. Phys. 2010, 82, 3045.

[16] X. Wan, A. M. Turner, A. Vishwanath, S. Y. Savrasov, Phys. Rev. B 2011, 83.

[17] Z. Wang, Y. Sun, X.-Q. Chen, C. Franchini, G. Xu, H. Weng, X. Dai, Z. Fang, Phys. Rev. B 2012, 85.

[18] Z. Zhu, J. E. Hoffman, Nature 2014, 513, 319.

[19] M. N. Ali, Q. Gibson, S. Jeon, B. B. Zhou, A. Yazdani, R. J. Cava, Inorg. Chem. 2014, 53, 4062.

[20] L. P. He, X. C. Hong, J. K. Dong, J. Pan, Z. Zhang, J. Zhang, S. Y. Li, Phys. Rev. Lett. 2014, 113, 246402.

[21] F. X. Xiu, L. A. He, Y. Wang, L. N. Cheng, L. T. Chang, M. R. Lang, G. A. Huang, X. F. Kou, Y. Zhou, X. W. Jiang, Z. G. Chen, J. Zou, A. Shailos, K. L. Wang, Nat. Nano. 2011, 6, 216.

[22] G. P. Mikitik, Y. V. Sharlai, Phys. Rev. Lett. 1999, 82, 2147.

[23] Y. B. Zhang, Y. W. Tan, H. L. Stormer, P. Kim, Nature 2005, 438, 201.

[24] J. Park, G. Lee, F. Wolff-Fabris, Y. Y. Koh, M. J. Eom, Y. K. Kim, M. A. Farhan, Y. J. Jo, C. Kim, J. H. Shim, J. S. Kim, Phys. Rev. Lett. 2011, 107, 126402.

[25] H. Murakawa, M. S. Bahramy, M. Tokunaga, Y. Kohama, C. Bell, Y. Kaneko, N. Nagaosa, H. Y. Hwang, Y. Tokura, Science 2013, 342, 1490.